\documentclass[aps,floatfix,eqsecnum,showpacs,preprint,tightenlines]{revtex4-1}
\usepackage{epsfig}
\usepackage{bm}
\usepackage{amsmath}
\usepackage{color}
\def\e{\kern+.5ex\lower.42ex\hbox{$\scriptstyle \iota$}\kern-1.10ex e}
\def\registered{{\ooalign{\hfil\raise .00ex\hbox{\scriptsize R}\hfil\crcr\mathhexbox20D}}}

\newcommand{\BA}[1]{\langle #1 \mid}

\newcommand{\KT}[1]{\mid #1 \rangle}

\begin{document}

\title{Muon capture on $^3$H}

\author{J. Golak}
\affiliation{M. Smoluchowski Institute of Physics, Jagiellonian University, PL-30384 Krak\'ow, Poland}
\author{R. Skibi{\'n}ski}
\affiliation{M. Smoluchowski Institute of Physics, Jagiellonian University, PL-30384 Krak\'ow, Poland}
\author{H. Wita{\l}a}
\affiliation{M. Smoluchowski Institute of Physics, Jagiellonian University, PL-30384 Krak\'ow, Poland}
\author{K. Topolnicki}
\affiliation{M. Smoluchowski Institute of Physics, Jagiellonian University, PL-30384 Krak\'ow, Poland}
\author{H. Kamada}
\affiliation{Department of Physics, Faculty of Engineering,
Kyushu Institute of Technology, Kitakyushu 804-8550, Japan}
\author{A. Nogga}
\affiliation{Institut f\"ur Kernphysik (Theorie),
             Institute for Advanced Simulation,
             J\"ulich Center for Hadron Physics and JARA - High Performance Computing, 
             Forschungszentrum J\"ulich,
             D-52425 J\"ulich, Germany}
\author{L.E. Marcucci}
\affiliation{Department of Physics, University of Pisa, IT-56127 Pisa, Italy
              and INFN-Pisa, IT-56127 Pisa, Italy}

\date{\today}

\begin{abstract}
The
$\mu^- + ^3\textrm{H} \rightarrow \nu_\mu + n + n + n$
capture reaction
is studied 
under full inclusion of final state interactions.
Predictions for the three-body break-up of $^3$H
are calculated with the AV18 potential, augmented by the Urbana~IX 
three-nucleon force. Our results are based on 
the single nucleon weak current operator comprising the dominant relativistic
corrections. 
This work is a natural extension of our investigations 
of the 
$\mu^- + ^3\textrm{He} \rightarrow \nu_\mu + ^3\textrm{H}$
$\mu^- + ^3\textrm{He} \rightarrow \nu_\mu + n + d$
and
$\mu^- + ^3\textrm{He} \rightarrow \nu_\mu + n + n + p$
capture reactions presented in Phys.\ Rev.\ C {\bf 90}, 024001 (2014).
\end{abstract}


\maketitle

\section{Introduction}
\label{section1}

Muon capture reactions on light nuclei have been studied intensively, both experimentally 
and theoretically, for many years. Earlier achievements were summarized in Refs.~\cite{Mea01,Gor04,Kam10}. 
More recent theoretical work focused
on the $ \mu^- + ^2\textrm{H} \rightarrow \nu_\mu + n + n $
and
$ \mu^- + ^3\textrm{He} \rightarrow \nu_\mu + ^3\textrm{H} $
reactions and was described in Refs.~\cite{prc83.014002,Mar12,Mar11b}.
The calculation of Ref.~\cite{prc83.014002}
was performed both in the phenomenological and the
``hybrid'' chiral effective field theory ($\chi$EFT) approach,
initiated in Ref.~\cite{Mar02}. It was based on 
Hamiltonians comprising
two-nucleon (2N)
as well as three-nucleon (3N) potentials.
The weak current operator included not only the single nucleon 
contribution but also
meson-exchange currents (MEC) as well as
currents arising from the $\Delta$-isobar excitation~\cite{prc63.015801}.
Later these two reactions were studied in a ``non-hybrid''
$\chi$EFT approach~\cite{Mar12b,Mar14}, where both potentials and currents are
derived consistently from $\chi$EFT. The results obtained within 
different approaches agree with each other and described 
the available experimental data well.

In Ref.~\cite{golak2014} we joined our expertise: 
from the momentum space treatment of 
electromagnetic processes \cite{physrep,romek2} 
and from the potential model approach developed in 
Ref.~\cite{prc83.014002}. 
We found that new results for the 
$ \mu^- + ^2\textrm{H} \rightarrow \nu_\mu + n + n $
and
$ \mu^- + ^3\textrm{He} \rightarrow \nu_\mu + ^3\textrm{H} $
reactions calculated in the momentum space were in good agreement
with those of Ref.~\cite{prc83.014002}, which had been obtained using 
the hyperspherical harmonics formalism. Thus we could make the first step
to establish a theoretical framework which can be extended to all 
the $A\leq 3$ muon capture reactions, including three-body 
break-up of the $A=3$ systems. By using the Faddeev equation
approach, we provided, for the 
first time, predictions for the total and differential capture rates of the
$\mu^- + ^3\textrm{He} \rightarrow \nu_\mu + n + d$ 
and 
$\mu^- + ^3\textrm{He} \rightarrow \nu_\mu + n + n + p$
break-up reactions,
calculated with the full inclusion of final state 2N and 3N interactions.
Although we incorporated selected MEC in the momentum space treatment 
of the 
$ \mu^- + ^2\textrm{H} \rightarrow \nu_\mu + n + n $
and
$ \mu^- + ^3\textrm{He} \rightarrow \nu_\mu + ^3\textrm{H} $
capture reactions, in the calculations of the break-up channels in muon 
capture on $^3$He we restricted ourselves to the single nucleon current, 
with the weak nucleon form factors from Ref.~\cite{She12}.

Muon capture on $^3$H has attracted less attention.
This reaction, with all uncharged particles in the final state, 
would be very difficult to measure because of the 
radioactivity of the target and due to the meso-molecular complications~\cite{Mea01}.
The 
$ \mu^- + ^3\textrm{H} \rightarrow \nu_\mu + n + n + n $
capture process presents, however, interesting features which make this process 
worth investigating: it allows one to study the neutron-neutron interaction
and the three-neutron force acting exclusively in the total isospin $T=3/2$
state. Besides, its study is the natural next step 
after the 
$ \mu^- + ^3\textrm{He} \rightarrow \nu_\mu + n + n + p $
reaction has been considered.
Theoretical studies of muon capture on $^3$H were started in the seventies of the 
20th century~\cite{phillips1975,torre1978,torre1979}. Those early calculations were 
performed predominantly in configuration space, using the 2N potential models available at that
time.  In Ref.~\cite{phillips1975} a separable potential of the Yamaguchi type
was employed in the calculations based on Amado's method~\cite{amado1963}
and thus the final state interaction (FSI) was taken into account. 
Actually that paper focused on the three different muon capture reactions 
in $^3$He and the information about the capture rate on $^3$H was extracted 
from the total isospin $T=3/2$ rate calculated for the three-body 
breakup of $^3$He.
The FSI effects and meson exchange currents were neglected in Ref.~\cite{torre1978}
but some observations about the reaction mechanism proved to be correct.
In particular the authors predicted that inclusion of FSI would lead to an 
enhancement of the muon capture rate.
A better calculation scheme was introduced in Ref.~\cite{torre1979}. 
The authors presented a general method 
to deal with transitions from a 3N bound state to scattering states
caused by a weakly acting Hamiltonian and applied it to muon capture on $^3$H.
They obtained results not only under plane wave (PW) approximation 
but also including 2N interactions (in the form of the supersoft-core nucleon-nucleon 
potential~\cite{detourreil1976}) in the three-neutron continuum. More recent 
theoretical investigations were conducted 
in Ref.~\cite{dzhibuti1984}. The authors 
used the method of hyperspherical functions in coordinate space 
and employed four different central potentials in their calculations. 
As in Ref.~\cite{torre1979}, they found FSI effects to be very important.
Their results were sensitive to the form of the 2N potential used 
in the calculations. Table~I in Ref.~\cite{dzhibuti1984}
nicely summarized all the early theoretical predictions.

In this paper we extend our investigations of the 
$\mu^- + ^3\textrm{He} \rightarrow \nu_\mu + n + d$
and
$\mu^- + ^3\textrm{He} \rightarrow \nu_\mu + n + n + p$
capture reactions presented in Ref.~\cite{golak2014} 
to describe also the
$\mu^- + ^3\textrm{H} \rightarrow \nu_\mu + n + n + n$
process. This reaction 
is studied 
under full inclusion of final state interactions,
employing the AV18 2N potential~\cite{av18} alone 
or together with the Urbana~IX 
3N force \cite{urbana}. Our results are based on 
the single nucleon weak current operator
including relativistic corrections~\cite{golak2014}.

The paper is organized in the following way:
In Sec.~\ref{section2} we briefly introduce the
elements of our formalism.
Our main results are shown in Sec.~\ref{section3},
where we discuss various predictions obtained with different 
dynamics for the
$\mu^- + ^3\textrm{H} \rightarrow \nu_\mu + n + n + n$ reaction.
We show our results for the differential and integrated rates
and compare them with earlier theoretical predictions.
To the best of our knowledge, we are for the first time able
to include final state interactions based on modern 2N and 3N 
forces in a way consistent with the bound state calculations.
In all cases we show separate results 
for the capture rates from the two hyperfine states, $F=0$ and $ F=1$, 
of the muon-tritium atom.
Finally, Sec.~\ref{section4} contains concluding remarks.

\section{Theoretical formalism}
\label{section2}

For the muon capture process one assumes that the initial state 
$ \KT{i\, } $ consists of the atomic $K$-shell muon wave function 
$ \KT{ \psi \, m_\mu \, } $ with the muon spin projection $m_\mu$ 
and the initial nucleus state with the three-momentum ${\bf P}_i$ 
(and the spin
projection $m_{i}$):
\begin{eqnarray}
\KT{i\, } = \KT{\psi \, m_\mu \, } \, \KT{\Psi_i \, {\bf P}_i \, m_{i} \, } \, .
\label{i}
\end{eqnarray}
In the final state, $\KT{f\, }$, one encounters
the muon neutrino 
(with the three-momentum ${\bf p}_{\nu}$ 
and the spin projection $m_{\nu}$), 
as well as the final nuclear state with the total 
three-momentum ${\bf P}_f$ and the set of spin projections $m_{f}$:
\begin{eqnarray}
\KT{f\, } = \KT{\nu_\mu \, {\bf p}_{\nu} \, m_\nu \, } \, 
\KT{\Psi_f \, {\bf P}_f \, m_{f} \, } \, .
\label{f}
\end{eqnarray}
The transition from the initial to the final state is driven by the 
Fermi form of the interaction Lagrangian (see for example Ref.~\cite{walecka})
and leads to a contraction of the leptonic (${\cal L}_\lambda$) and nuclear 
(${\cal N}^\lambda$) parts 
in the $S$-matrix element, $S_{fi}$ \cite{romek}:
\begin{eqnarray}
S_{fi}= i ( 2 \pi )^4 \, \delta^4 \left( P^\prime - P \right)\, 
\frac {G}{\sqrt{2}} \, {\cal L}_\lambda \, {\cal N}^\lambda \, ,
\label{sfi}
\end{eqnarray}
where $G= 1.14939 \times 10^{-5} \, {\rm GeV}^{-2} $ is the Fermi constant
(taken from Ref.~\cite{prc83.014002}), 
and $P$ ($P^\prime$) is the 
total initial (final) four-momentum. 
The well known leptonic matrix element 
\begin{eqnarray}
{\cal L}_\lambda = \frac 1{ \left( 2 \pi \, \right)^3 } \, 
\bar{u} ( {\bf p}_{\nu} , m_{\nu} ) \gamma_\lambda ( 1- \gamma_5 ) 
u ( {\bf p}_{\mu} , m_{\mu} )
\, \equiv \,
\frac 1{ \left( 2 \pi \, \right)^3 } \, L_\lambda
\label{llambda}
\end{eqnarray}
is given in terms of the Dirac matrices and spinors.

The nuclear part is the essential ingredient of the formalism
and is written as~\cite{physrep,romek}
\begin{eqnarray}
{\cal N}^\lambda = \frac 1{ \left( 2 \pi \, \right)^3 } \, 
\BA{\Psi_f \, {\bf P}_f \, m_{f} \, } \, 
j_w^\lambda
\, \KT{\Psi_i \, {\bf P}_i \, m_{i} \, } 
\, \equiv \,
\frac 1{ \left( 2 \pi \, \right)^3 } \, N^\lambda \, .
\label{nlambda}
\end{eqnarray}
It is a matrix element of the nuclear weak current operator 
$j_w^\lambda$ between the initial and final nuclear states. 
In this paper we omit many-nucleon contributions to $j_w^\lambda$
and restrict ourselves to two forms of the 
single nucleon current operator.
The first one, $j_w^\lambda= {j}^\lambda_{\text{NR}}$, is strictly nonrelativistic, with 
the following momentum-space matrix elements of its time and space 
components~\cite{romek}:
\begin{eqnarray}
\BA{{\bf p}^{\, \prime}} {j}^{0}_{\text{NR}}  \KT{{\bf p}} = \left( g_{1}^{V} + g_{1}^{A} \frac{{{\bm \sigma}} 
\cdot \left( {\bf p} + {\bf p}^{\, \prime} \right)}{2 M}  \right) {\tau}_{-}
\label{jnr01}
\end{eqnarray}
and
\begin{eqnarray}
&&\BA{{\bf p}^{\, \prime}} {{\bf j}}_{\text{NR}}  \KT{{\bf p}\, } 
= \nonumber \\
&&\bigg( g_{1}^{V} \frac{{\bf p} + {\bf p}^{\, \prime}}{2 M} - \frac{1}{2 M} \left(g_{1}^{V} - 2 M g_{2}^{V} \, \right) 
i \, {{\bm \sigma}} \times \left( {\bf p} - {\bf p}^{\, \prime}  \, \right) \nonumber \\
&&+ g_{1}^{A} {{\bm \sigma}} + g_{2}^{A} \left( {\bf p} - {\bf p}^{\, \prime} \, \right) \frac{{{\bm \sigma}} \cdot 
\left({\bf p} - {\bf p}^{\, \prime} \, \right)}{2 M} \bigg) {\tau}_{-} \, ,
\label{jnrvec1}
\end{eqnarray}
where 
$M$ is the mean value of the proton ($M_p$)
and neutron ($M_n$) masses, $M \equiv \frac12 \left( M_p + M_n \, \right) $,
${\tau}_{-} \equiv(\tau_x -{\rm i} \tau_y)/2$ is the isospin lowering operator,
${{\bm \sigma}}$ is a vector of Pauli spin matrices
and $ {\bf p}$ ($ {\bf p}^{\, \prime}$) is the initial (final) nucleon momentum.
Here we keep only terms up to $1/M$.

The second form of $j_w^\lambda$, ${j}^\lambda_{\text{NR+RC}}$, 
contains relativistic $1/M^2$ corrections, which 
leads to additional terms in the corresponding 
matrix elements~\cite{golak2014}:
\begin{eqnarray}
&&\BA{{\bf p}^{\, \prime}} {j}^{0}_{\text{NR+RC}}  \KT{{\bf p}} = \nonumber \\
&&\bigg( g_{1}^{V} - (g_{1}^{V} - 4 M g_{2}^{V}) \frac{\left({\bf p}^{\, \prime} - {\bf p} \, \right)^{2}}{8 M^{2}} + 
\left(g_{1}^{V} - 4 M g_{2}^{V} \, \right) \, i \, \frac{\left( {\bf p}^{\, \prime} \times {\bf p} \, \right) \cdot 
{{\bm \sigma}}}{4 M^{2}} \nonumber \\
&&+ g_{1}^{A} \frac{{{\bm \sigma}} \cdot \left( {\bf p} + {\bf p}^{\, \prime} \, \right)}{2 M} + g_{2}^{A} 
\frac{\left( {\bf p}^{\, \prime\, 2} - {\bf p}^{2} \, \right)}{4 M^{2}} {{\bm \sigma}} \cdot \left({\bf p}^{\, \prime} - 
{\bf p} \, \right)  \bigg) {\tau}_{-} 
\label{1nr+rc01}
\end{eqnarray}
and
\begin{eqnarray}
&&\BA{{\bf p}^{\, \prime}} {{\bf j}}_{\text{NR+RC}}  \KT{{\bf p}} = \nonumber \\
&&\bigg( g_{1}^{V} \frac{{\bf p} + {\bf p^{\, \prime}}}{2 M} - \frac{1}{2 M} \left(g_{1}^{V} - 2 M g_{2}^{V} \, \right) 
\, i \, {{\bm \sigma}} \times \left( {\bf p} - {\bf p}^{\, \prime} \, \right) \nonumber \\
&&+ g_{1}^{A} \left(1 - \frac{\left({\bf p} + {\bf p}^{\, \prime} \, \right)^{2}}{8 M^{2}} \, \right) {{\bm \sigma}} + 
\nonumber \\
&&+ \frac{g_{1}^{A}}{4 M^{2}} \big[ \left({\bf p} \cdot {{\bm \sigma}} \, \right) {\bf p}^{\, \prime} + \left({\bf p}^{\, \prime} 
\cdot {{\bm \sigma}} \, \right) {\bf p} + \, i \, \left( {\bf p} \times 
{\bf p}^{\, \prime} \,  \right) \big]) \nonumber \\
&&+g_{2}^{A} \left( {\bf p} - {\bf p}^{\, \prime} \, \right) \frac{{{\bm \sigma}} \cdot \left({\bf p} - {\bf p}^{\, \prime} \, \right)}{2 M} 
\bigg) {\tau}_{-} \ .
\label{1nr+rcvec1}
\end{eqnarray}
This form of the nuclear weak current operator is 
very close to the one used in Ref.~\cite{prc83.014002}, see Ref.~\cite{golak2014} for details.
Note that the weak nucleon form factors 
$g_1^V$, $g_2^V$, $g_1^A$ and $g_2^A$
are usually expressed in terms of the 
isovector components of the electric ($G_E^V$) 
and magnetic ($G_M^V$) Sachs form factors as well 
as the axial ($G_{A}$) and pseudoscalar ($G_{P}$) form factors:
\begin{eqnarray}
G_E^V&=&g_1^V  \, , \label{gev} \\
G_M^V&=&g_1^V-2M g_2^V \, , \label{gmv} \\
G_{A} &=& -g_{1}^{A} \, , \label{ga}\\
G_{P} &=& -g_{2}^{A} m_{\mu} \, . \label{gp}
\end{eqnarray}
As in Ref.~\cite{golak2014}, in this paper 
we also employ the form factors from Ref.~\cite{She12}.

The essential part of the decay rate formula stems from the contraction of the 
leptonic and nuclear matrix elements.  Note that, contrary to what was erroneously stated in 
Ref.~\cite{golak2014}, we use indeed the same notation as
Bjorken and Drell \cite{bjodrell} but with the different spinor
normalization. To be explicit,
we use the following definitions:
\begin{eqnarray}
u ( {\bf p} , s ) \equiv \, \sqrt{ \frac{E + m}{2 \, E}  } \,
 \left( 
\begin{array}{c}
\chi_s\\ [2pt]
\frac{  {\bf p} \cdot {\bf \sigma} }{E + m} \, \chi_s 
\end{array}
 \right) \, ,
\label{spinor1}
\end{eqnarray}
which means that 
$u^\dagger \, u = 1$ 
and 
$\bar{u} \, u = \frac{m}{E} $,
where $m$ is the particle mass and $E \equiv \sqrt{ m^2 + {\bf p}^2 \, }$.
We assume of course that the two-component spinor $\chi_s$ is normalized to yield $ \chi_s^\dagger \, \chi_s = 1$.

Since we deal with a specific muon capture reaction,
we switch from the general notation to the one where
the relevant spin projections are given explicitly:
\begin{eqnarray}
L_\lambda &\equiv& L_\lambda ( m_\nu , m_\mu \, ) \, , \nonumber \\
N^\lambda &\equiv& N^\lambda ( m_1, m_2, m_3, m_{^3H} \, ) 
\end{eqnarray}
and use in the following the minus one spherical component of ${\bf N}$: 
$N_{-1} = \frac1{\sqrt{2}} \left( N_x - i N_y \,  \right) $.
With these definitions, assuming additionally that
$ {\hat {\bf p}}_\nu = - {\hat {\bf z}} $ and that the initial muon is at rest, 
we easily evaluate for the unpolarized case
\begin{eqnarray}
\left| {\cal T} \right|^2 &\equiv& 
\frac14 \, \sum\limits_{m_{^3H},m_\mu} \sum\limits_{m_1, m_2, m_3, m_\nu } 
\left| L_\lambda ( m_\nu , m_\mu \, ) N^\lambda (m_1, m_2, m_3, m_{^3H} \, ) \, \right|^2  \nonumber \\
&=&\frac12 \sum\limits_{m_{^3H}} \sum\limits_{m_1, m_2, m_3 } 
\Big( 
\left| N^0 (m_1, m_2, m_3, m_{^3H} \, ) \, \right |^2 \, + \, 
\left| N_z (m_1, m_2, m_3, m_{^3H} \, ) \, \right |^2 
\nonumber \\
&+&2 \left| N_{-1} (m_1, m_2, m_3, m_{^3H} \, ) \, \right |^2 
\nonumber \\
&+&2 {\rm Re} \left( N^0 (m_1, m_2, m_3, m_{^3H} \, )  \left( N_z (m_1, m_2, m_3, m_{^3H} \, ) \right)^*   \right ) \, \Big)  .
\label{contract2}
\end{eqnarray}

This form is not appropriate when we want to separately calculate capture rates 
from two hyperfine states $F=0$ or $ F=1$ of the muon-tritium atom.
In such a case we introduce the coupling between the triton and muon spin
via standard Clebsch-Gordan coefficients $c ( \frac12 , \frac12 , F ; m_\mu , m_{^3H} , m_F \, ) $ 
and obtain
\begin{eqnarray}
\left| {\cal T} \right|^2_F &\equiv& 
\frac1{2F + 1} \sum\limits_{m_F} \sum\limits_{m_1, m_2 , m_3, m_\nu}  \nonumber \\
&&\Big| 
\sum\limits_{m_\mu , m_{^3H}} \, 
c ( \frac12 , \frac12 , F ; m_\mu , m_{^3H} , m_F \, ) \, 
L_\lambda ( m_\nu , m_\mu \, ) N^\lambda (m_1, m_2, m_3, m_{^3H} \, ) \, \Big|^2 \, .
\label{contract3}
\end{eqnarray}
The explicit formulas for $F=0$ and $F=1$ read
\begin{eqnarray}
\left| {\cal T} \right|^2_{F=0} &=& \sum\limits_{m_1, m_2 , m_3}  
\Big| N^0 (m_1, m_2, m_3, m_{^3H}=-\frac12 \, ) \nonumber \\
&-&\sqrt{2} \, N_{-1} (m_1, m_2, m_3, m_{^3H}=\frac12 \, ) \, + \, 
N_z (m_1, m_2, m_3, m_{^3H}=-\frac12 \, ) \, 
\Big|^2 \, . 
\label{contract4}
\end{eqnarray}
and
\begin{eqnarray}
\left| {\cal T} \right|^2_{F=1} &=& \frac23 \, \sum\limits_{m_1, m_2 , m_3}  
\Big( \,
\Big| N^0 (m_1, m_2, m_3, m_{^3H}=\frac12 \, ) \, \Big|^2    \nonumber \\
&+&2\, \Big| N_{-1} (m_1, m_2, m_3, m_{^3H}=-\frac12 \, ) \, \Big|^2  \nonumber \\
&+&\Big| N_{-1} (m_1, m_2, m_3, m_{^3H}=\frac12 \, ) \, \Big|^2    \nonumber \\
&+&\frac12 \, \Big| N^0 (m_1, m_2, m_3, m_{^3H}=-\frac12 \, ) \, + \,
                 N_z (m_1, m_2, m_3, m_{^3H}=-\frac12 \, ) \, \Big|^2  \nonumber \\
&+&\Big| N_z (m_1, m_2, m_3, m_{^3H}=\frac12 \, ) \, \Big|^2  \nonumber \\
&+&2 {\rm Re} \left( N^0 (m_1, m_2, m_3, m_{^3H}=\frac12 \, )  \left( N_z (m_1, m_2, m_3, m_{^3H}=\frac12 \, ) \right)^*   \right)  \nonumber \\
&+&\sqrt{2} {\rm Re} \left( N^0 (m_1, m_2, m_3, m_{^3H}=-\frac12 \, )  \left( N_{-1} (m_1, m_2, m_3, m_{^3H}=\frac12 \, ) \right)^*   \right)  \nonumber \\
&+&\sqrt{2} {\rm Re} \left( N_z (m_1, m_2, m_3, m_{^3H}=-\frac12 \, )  \left( N_{-1} (m_1, m_2, m_3, m_{^3H}=\frac12 \, ) \right)^*   \right)  \, 
\, \Big) \,  .
\label{contract5}
\end{eqnarray}
These three quantities, $\left| {\cal T} \right|^2$, $\left| {\cal T} \right|^2_{F=0} $
and $\left| {\cal T} \right|^2_{F=1} $ are not independent but obey the obvious relation
\begin{eqnarray}
\left| {\cal T} \right|^2 \, = \, \frac14 \left| {\cal T} \right|^2_{F=0} \, + \,
\frac34 \left| {\cal T} \right|^2_{F=1} \, .
\label{notindpd}
\end{eqnarray}


The crucial matrix elements
\begin{eqnarray}  
N^\lambda (m_1, m_2 , m_3 , m_{^3\textrm{H}}  \, )  \, \equiv \, 
\BA{\Psi_{nnn}^{(-)}  \, {\bf P}_f=-{\bf p}_\nu  \, m_1 \, m_2 \, m_3} \, 
j_w^\lambda
\, \KT{\Psi_{^3\textrm{H}} \, {\bf P}_i=0 \, m_{^3\textrm{H}} \, } 
\label{nnnn}
\end{eqnarray}  
are calculated in two steps~\cite{physrep,romek2}.
First we solve a Faddeev-like equation 
for the auxiliary state $ \KT{ U^\lambda \, } $ for each considered 
neutrino energy:
\begin{eqnarray} 
\KT{ U^\lambda \, } = \Big[ t G_0 \, + \, \frac12 ( 1 + P ) V_4^{(1)}  G_0  ( 1 + t G_0 \, ) \, \Big] ( 1 + P ) j_w^\lambda \KT{\Psi_{^3\textrm{H}} \, } 
\nonumber \\  
+ \  \Big[ t G_0 P \, + \, \frac12 ( 1 + P ) V_4^{(1)}  G_0  ( 1 + t G_0 P \, ) \, \Big] \KT{ U^\lambda \, }  \, ,
\label{u}
\end{eqnarray}  
where $ V_4^{(1)} $ is a part of the 3N force symmetrical under the exchange of nucleon 2 and ~3,
$G_0$ is the free 3N propagator and $t$ is the 2N $t$-operator
acting in the $(2,3)$ subspace. Further $P$ is the permutation operator built from the 
transpositions $P_{ij}$ exchanging nucleons $i$ and $j$:  
\begin{eqnarray} 
P = P_{12} P_{23} + P_{13} P_{23} \, .
\label{p}
\end{eqnarray}  
In the second step the nuclear matrix elements are calculated 
by quadratures:
\begin{eqnarray}  
N^\lambda (m_1, m_2 , m_3 , m_{^3\textrm{H}}  \, )  &=&
\BA{\phi_{nnn} \, {\bf p} \, {\bf q} \, m_1 \, m_2 \, m_3 \, } \, ( 1 + P ) j_w^\lambda \KT{\Psi_{^3\textrm{H}} \, }
\nonumber \\
&+&
\BA{\phi_{nnn} \, {\bf p} \, {\bf q} \, m_1 \, m_2 \, m_3 \, } \, t G_0  ( 1 + P ) j_w^\lambda \KT{\Psi_{^3\textrm{H}} \, }
\nonumber \\
&+&
\BA{\phi_{nnn} \, {\bf p} \, {\bf q} \, m_1 \, m_2 \, m_3 \, } \,  P \KT{ U^\lambda \, }
\nonumber \\
&+&
\BA{\phi_{nnn} \, {\bf p} \, {\bf q} \, m_1 \, m_2 \, m_3 \, } \,  t G_0 P \KT{ U^\lambda \, }  \, .
\label{nnnn2}
\end{eqnarray}  
Here 
$\KT{\phi_{nnn} \, {\bf p} \, {\bf q} \, m_1 \, m_2 \, m_3 \, }$
is a product state of Jacobi momenta $ {\bf p} $ and ${\bf q} $
describing two free motions in the three-neutron system 
\begin{eqnarray}
{\bf p}  &\equiv& \frac12 \left(  {\bf p}_2 - {\bf p}_3 \,  \right) \, , \nonumber \\
{\bf q}  &\equiv& \frac23 \left(  {\bf p}_1 - \frac12 \left(  {\bf p}_2 +  {\bf p}_3 \,  \right) \,  \right) \, 
= \, {\bf p}_1 + \frac13 {\bf p}_\nu \, .
\label{pq}
\end{eqnarray}
Equations (\ref{u}) and (\ref{nnnn2}) simplify significantly,
when $ V_4^{(1)} = 0$ \cite{romek2}. In this case one obtains 
\begin{eqnarray} 
\KT{ U^\lambda \, } = t G_0 \, ( 1 + P ) j_w^\lambda \KT{\Psi_{^3\textrm{H}} \, } 
\, + \,  t G_0 P \, \KT{ U^\lambda \, }
\label{usimple}
\end{eqnarray}  
and
\begin{eqnarray}  
N^\lambda (m_1, m_2 , m_3 , m_{^3\textrm{H}}  \, )  &=&
\BA{\phi_{nnn} \, {\bf p} \, {\bf q} \, m_1 \, m_2 \, m_3 \, } \, ( 1 + P ) j_w^\lambda \KT{\Psi_{^3\textrm{H}} \, }
\nonumber \\
&+&
\BA{\phi_{nnn} \, {\bf p} \, {\bf q} \, m_1 \, m_2 \, m_3 \, } \, ( 1 +  P ) \KT{ U^\lambda \, } \, .
\label{nnnn2simple}
\end{eqnarray}

Taking all factors into account, using the rotational symmetries of the unpolarized case 
and evaluating the phase space factor 
in terms of the relative Jacobi momenta ${\bf p}$ and ${\bf q}$, 
we arrive at the final 
expression for the total capture rate
for the $ \mu^- + ^3\textrm{H} \rightarrow \nu_\mu + n + n + n $ reaction: 
\begin{eqnarray}
&&\Gamma = \frac32 G^2 \frac1{( 2 \pi )^2 } \, 
{\cal R} \, \frac { \left( M^\prime_{^3\textrm{H}} \alpha \, \right)^3 } {\pi  } \, 4 \pi \, 
\nonumber \\
&&\int\limits_0^{E_\nu^{max,nnn}} \, dE_\nu E_\nu^2  \,
\frac13 \, 
\int\limits_{0}^{\pi} d \theta_{q} \sin \theta_{q} \, 2 \pi \, 
\int\limits_{0}^{\pi} d \theta_{p} \sin \theta_{p} \, \int\limits_{0}^{2 \pi} d \phi_{p} \, 
\int\limits_0^{p^{max}} \, dp p^2  \, \frac23 M q  \, \left| {\cal T} \right|^2 \, ,
\label{gnnn}
\end{eqnarray}  
where the factor 
$\frac { \left( M^\prime_{^3\textrm{H}} \alpha \, \right)^3 } {\pi  } $
stems from the 
$K$-shell atomic wave function, 
$M^\prime_{^3\textrm{H}} = \frac{ M_{^3\textrm{H}} \, M_\mu } { M_{^3\textrm{H}} + M_\mu}$
and  $ \alpha \approx \frac 1{137} $ is the fine structure constant. 
The value of $q \equiv \left| {\bf q} \right| $ is defined through 
Eq.~(\ref{epq}) below.
The additional factor ${\cal R}$ can account for the finite volume 
of the $^3$H charge but we take ${\cal R}= 1$ in the present calculations.
Note that the current operator of nucleon~1 is used when evaluating $\left| {\cal T} \right|^2$. 

In order to fix the upper limit of the integration over $p$ in (\ref{gnnn}), we express
the energy conservation in terms of the Jacobi momenta:
\begin{eqnarray}
M_\mu + M_{^3\textrm{H}} \approx E_\nu + 3 M
+ \frac{ {\bf p}^{\ 2}} {M} 
+ \frac34 \frac{ {\bf q}^{\ 2}} {M} 
+ \frac16 \frac{E_\nu^2 } {M} \, .
\label{epq}
\end{eqnarray}  

Like for the $\mu^- + ^2\textrm{H} \rightarrow \nu_\mu + n + n $ reaction studied
in Ref.~\cite{golak2014}, we can consider the hyperfine states in the muon-tritium atom,
replacing  
$ \left| {\cal T} \right|^2 $ 
by 
$ \left| {\cal T} \right|^2_{F=0} $ or 
$ \left| {\cal T} \right|^2_{F=1} $.

\section{Results for the $\mu^- + ^3\textrm{H} \rightarrow \nu_\mu + n + n + n $ reaction}
\label{section3}

We start with
the kinematics of the 
$\mu^-+ ^3\textrm{H} \rightarrow \nu_\mu + n + n + n $ 
reaction, which is formulated exactly in the same way as in \cite{golak2014}
for the $\mu^-+ ^3\textrm{He} \rightarrow \nu_\mu + n + n + p$
process.
The relativistic ({\it rel}) and non-relativistic ({\it nrl})
maximal neutrino energy
for this three-body capture of the muon atom 
is evaluated as

\begin{eqnarray}
\left( E_\nu^{max,nnn} \, \right)^{rel} &=&
\frac{{M_{^3\textrm{H}}}^2+2 {M_{^3\textrm{H}}}
   {M_\mu}+{M_\mu}^2 - 9{M_n}^2}{2
   ({M_{^3\textrm{H}}}+{M_\mu})} \, = 94.3078\ {\rm MeV} \, ,
\label{Enumaxrelnnn} \\
\left( E_\nu^{max,nnn} \, \right)^{nrl} &=&
\sqrt{
 3 M_n \, ( 2 M_{^3\textrm{H}} + 2 M_\mu - 3 M_n ) } - 3 {M_n} \, = \, 94.3073\ {\rm MeV} \, .
\label{Enumaxnrlnnn}
\end{eqnarray}

\begin{figure}
\begin{center}
\includegraphics[width=9cm]{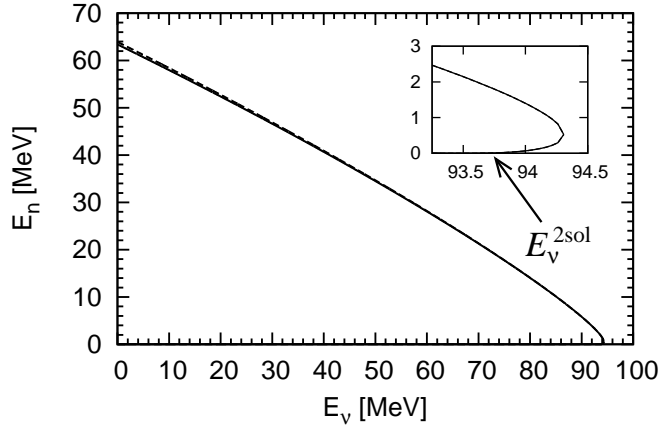}
\end{center}
\caption{The kinematically allowed region in the
$E_\nu - E_n$ plane calculated relativistically
(solid curve) and
nonrelativistically (dashed curve) for the
$\mu^-+ ^3\textrm{H} \rightarrow \nu_\mu + n + n + n $ process. \label{enu_en}}
\end{figure}

The kinematically allowed region 
in the $E_\nu - E_n$ plane for the 
break-up of $^3$H is shown in Fig.~\ref{enu_en}.
We show the curves based on the relativistic and nonrelativistic kinematics.
They essentially overlap except for very small neutrino energies.
Up to a certain $E_\nu$ value, which we denote by
$E_\nu^{2sol}$, the minimal neutron kinetic energy is zero. 
The minimal neutron kinetic energy is greater than zero
for $ E_\nu > E_\nu^{2sol}$. Even this very detailed feature 
of the kinematical domain can be calculated 
nonrelativistically with high accuracy (see also the inset in Fig.~\ref{enu_en}).
The  values of $E_\nu^{2sol}$
based on the relativistic kinematics, 
\begin{equation}
\left( E_\nu^{2sol} \, \right)^{rel} =
\frac{
( M_{^3\textrm{H}} + M_\mu ) ( M_{^3\textrm{H}} + M_\mu - 2 M_n ) - 3 {M_n}^2
}{2 \,  ({M_{^3\textrm{H}}}+{M_\mu}-{M_n})}
\label{Enu2solrel}
\end{equation}
and nonrelativistic kinematics, 
\begin{equation}
\left( E_\nu^{2sol} \, \right)^{nrl} =
2 \left(\sqrt{{M_{^3\textrm{H}}}
   {M_n}+{M_\mu}
   {M_n} - 2 {M_n}^2 }-{M_n}\right) \, ,
\label{Enu2solnrl}
\end{equation}
yield very similar numerical values, 
93.5574 MeV
and
93.5561 MeV, respectively.
This supports the notion 
that predictions based on the nonrelativistic 
3N dynamics should be valid for the considered capture process. 

Solutions of Eqs.~(\ref{u}) and (\ref{usimple})
as well as the evaluations of the nuclear matrix elements 
in Eqs.~(\ref{nnnn2}) and
in Eqs.~(\ref{nnnn2simple}) 
are obtained using partial wave decomposition (PWD).
We employ our standard 3N basis $\mid  p q \bar{\alpha }\, J m_J ; T m_T \, \rangle $
\cite{physrep}, where $p$ and $q$ are magnitudes of the relative Jacobi
momenta and $ \bar{\alpha}$ is a set of discrete quantum numbers.
Note that the $\mid  p q \bar{\alpha } \, J m_J ; T m_T \, \rangle $ states are 
already antisymmetrized in the $(2,3)$ subsystem. 
The initial 3N bound state is therefore represented as
\begin{eqnarray}  
\mid \Psi_{^3\textrm{H}} \, m_{^3\textrm{H}} \, \rangle  =
\sum\limits_{\bar{\alpha}_b} \,
\int dp p^2 \int dq q^2 \, 
\Big| p q \bar{\alpha}_b \, \frac12 \, m_{^3\textrm{H}} \, ;
\frac12 , -\frac12 \Big\rangle \, 
\phi_{\bar{\alpha}_b} \left( p , q \right) \, .
\label{pwdj13}
\end{eqnarray}  
In our calculation of the 3N bound state we use 34 (20) points 
for integration over $p$
($q$), and 34 partial wave states corresponding to the 2N subsystem 
total angular momentum $j \le 4$.   

In Ref.~\cite{golak2014} we checked that 
it is sufficient to perform calculations in 3N continuum with $j \le 3$.
The convergence with respect to the total 3N angular momentum $J$
is also very rapid and in the present 
calculations we include all the 3N partial wave states up to 
$J_{max}= \frac92$.
The first building block in our scheme requires 
PWD of the single nucleon current operator:
\begin{eqnarray}  
\langle p q \bar{\alpha} J m_J ; T m_T \, {\bf P}_f \mid j_w (1) \mid \Psi_{^3\textrm{H}} \, {\bf P}_i=0 \, m_{^3\textrm{H}} \, \rangle  \, .
\end{eqnarray} 
This step is described in detail in Ref.~\cite{golak2014}.

We refer the reader to Ref.~\cite{physrep} for the detailed definitions of 
various 3N dynamics. Here we only note that our plane wave (PW) 
predictions shown in the following for the sake of comparison with 
the earlier calculations are obtained with 
\begin{eqnarray}  
N^\lambda (m_1, m_2 , m_3 , m_{^3\textrm{H}}  \, )  =
\BA{\phi_{nnn} \, {\bf p} \, {\bf q} \, m_1 \, m_2 \, m_3 \, } \, ( 1 + P ) j_w^\lambda \KT{\Psi_{^3\textrm{H}} \, } \, .
\label{nnn3pw}
\end{eqnarray}

\begin{table}
\caption{Capture rates ($\Gamma$) for the 
$\mu^- + ^3\textrm{H} \rightarrow \nu_\mu + n + n + n $  
process calculated with the AV18~\cite{av18} nucleon-nucleon potential 
and the non-relativistic single nucleon current operator (Full 2NF),  
including relativistic corrections in the single nucleon current operator (Full 2NF RC) 
and additionally employing the Urbana~IX~\cite{urbana} 3N force (Full 2NF+3NF RC).
Also predictions obtained using the plane wave approximation 
are shown in brackets. Earlier theoretical 
predictions, obtained without a 3N force, are displayed in the same way.
For the information about the various 2N forces
(YAM, RSC, SSC, V, EH, S1, S2) used in 
Refs.~\cite{phillips1975,torre1978,torre1979,dzhibuti1984}
we refer the reader to the corresponding papers.
\label{tabbr}}
\begin{center}
\begin{tabular}{llcc}
\hline\hline
& \multicolumn{3}{c}{capture rate $\Gamma$ in s$^{-1}$} \\ \cline{2-4}
& $F=0$ & $F=1$ & total \\
\hline
Full 2NF        & 138.1 (100.0)   & 3.97 (2.97) & 37.5 (27.2) \\
Full 2NF RC     & 133.6 (97.0)   & 4.21 (3.12) & 36.5 (26.6) \\
Full 2NF+3NF RC & 118.7   & 3.92 & 32.6 \\
\hline
earlier theoretical predictions:  &     &  &              \\
Ref.~\cite{phillips1975} YAM   &     &  &  9.5 (6.1)  \\
Ref.~\cite{torre1978} RSC  &     &  &  (23.6)  \\
~~~~~~~~~~~ RSC RC &     &  &  (28.2)  \\
~~~~~~~~~~~ SSC  &     &  &  (23.0)  \\
~~~~~~~~~~~ SSC RC &     &  &  (27.6)  \\
Ref.~\cite{torre1979} SSC  &  122.8 (90.6)  & 3.58 (2.69)  &  33.4 (24.7) \\
~~~~~~~~~~~ SSC RC &  137.5 (102.0)  & 3.66 (2.78)   & 37.1 (27.6) \\
Ref.~\cite{dzhibuti1984} V  &    &   & 35.7 (22.3)  \\
~~~~~~~~~~~ EH   &    &   &  29.9 (19.7)  \\
~~~~~~~~~~~ S1   &    &   &  33.1 (20.8)  \\
~~~~~~~~~~~ S2   &    &   &  35.5 (21.9)  \\

\end{tabular}
\end{center}
\end{table}

\begin{figure}
\begin{center}
\includegraphics[width=9cm]{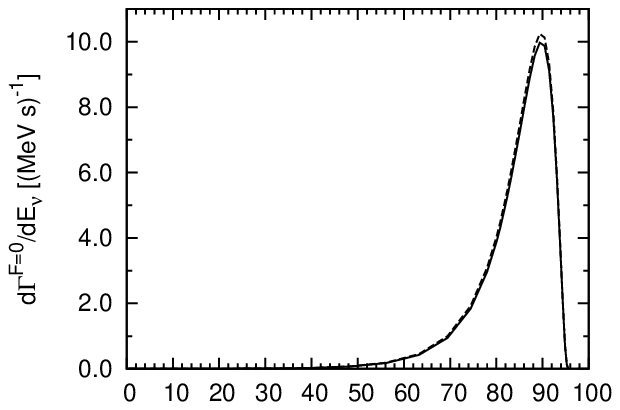}
\end{center}
\begin{center}
\includegraphics[width=9cm]{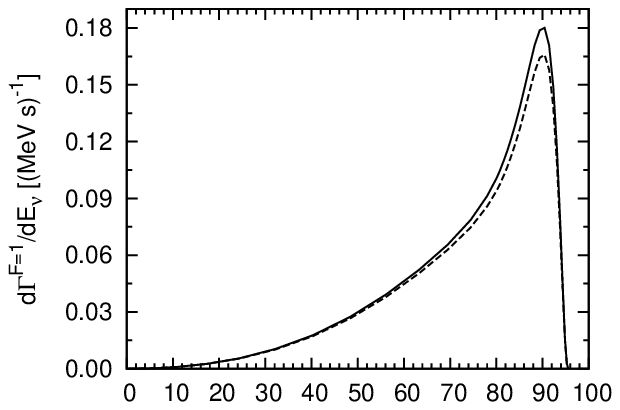}
\end{center}
\begin{center}
\includegraphics[width=9cm]{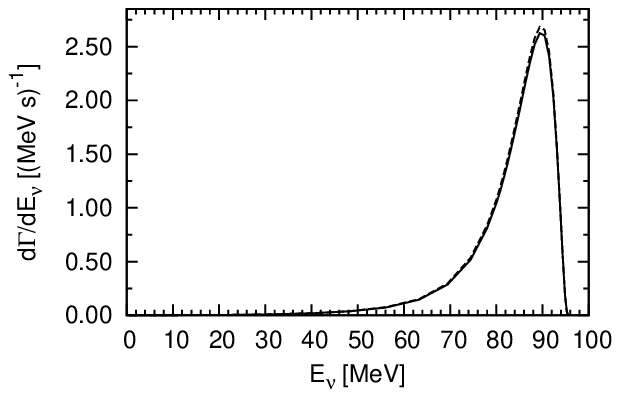}
\end{center}
\caption{The differential capture rates 
    ($F=0$) $ {d\Gamma }^{F=0}/ {dE_\nu} $ (top row),
    ($F=1$) $ {d\Gamma }^{F=1}/ {dE_\nu}   $ (middle row)
and (total) $ {d\Gamma }/ {dE_\nu} $ (bottom row),   
for the $ \mu^- + ^{3}\textrm{H} \rightarrow \nu_\mu + n + n + n$
process calculated with the single nucleon current operator without 
(dashed line) and with (solid line) relativistic corrections.
The calculations are based on the AV18 nucleon-nucleon potential \cite{av18} only.
\label{fig1}}
\end{figure}

\begin{figure}
\begin{center}
\includegraphics[width=9cm]{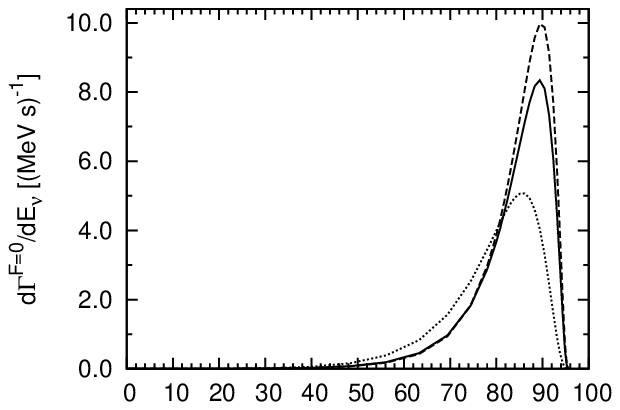}
\end{center}
\begin{center}
\includegraphics[width=9cm]{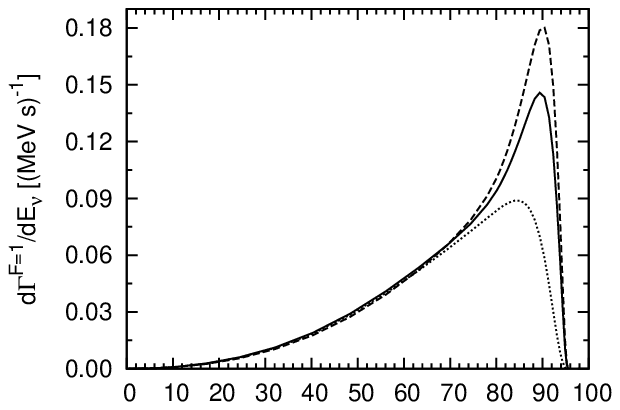}
\end{center}
\begin{center}
\includegraphics[width=9cm]{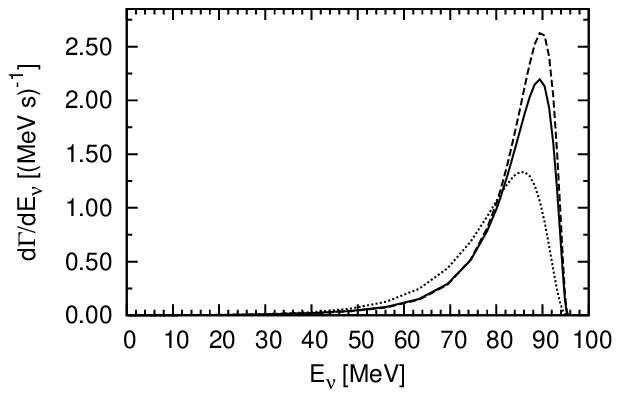}
\end{center}
\caption{The differential capture rates
    ($F=0$) $ {d\Gamma }^{F=0}/ {dE_\nu} $ (top row),
    ($F=1$) $ {d\Gamma }^{F=1}/ {dE_\nu}   $ (middle row)
and (total) $ {d\Gamma }/ {dE_\nu} $ (bottom row),
for the $ \mu^- + ^{3}\textrm{H} \rightarrow \nu_\mu + n + n + n$
process calculated with the single nucleon current operator
including relativistic corrections and
different types of 3N dynamics:
(symmetrized) plane wave (dotted curve),
with total omission of the 3N force
(dashed curve)
and with consistent inclusion 
of the 2N and 3N forces (solid curve).
The calculations are based on the AV18 nucleon-nucleon potential \cite{av18}
and the Urbana~IX 3N force \cite{urbana}.
\label{fig2}}
\end{figure}

\begin{figure}
\begin{center}
\includegraphics[width=9cm]{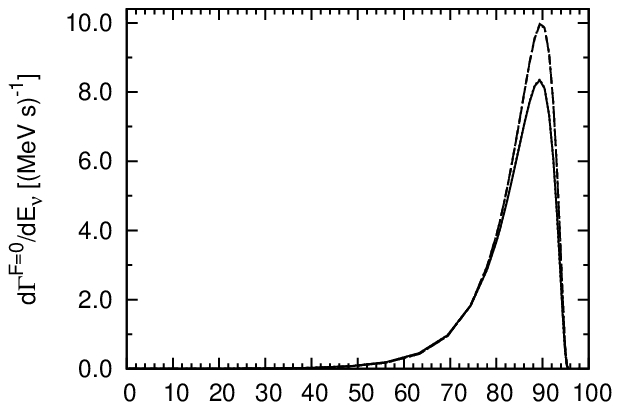}
\end{center}
\begin{center}
\includegraphics[width=9cm]{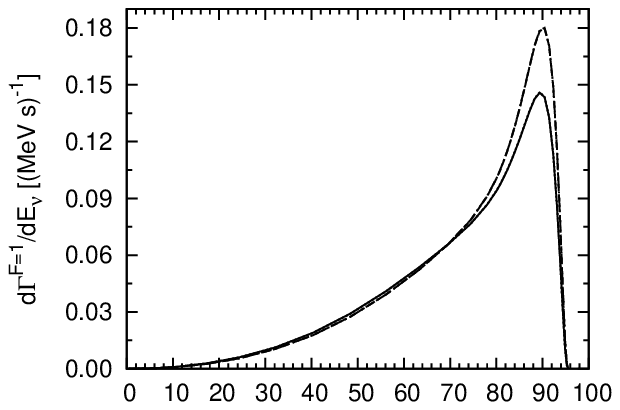}
\end{center}
\begin{center}
\includegraphics[width=9cm]{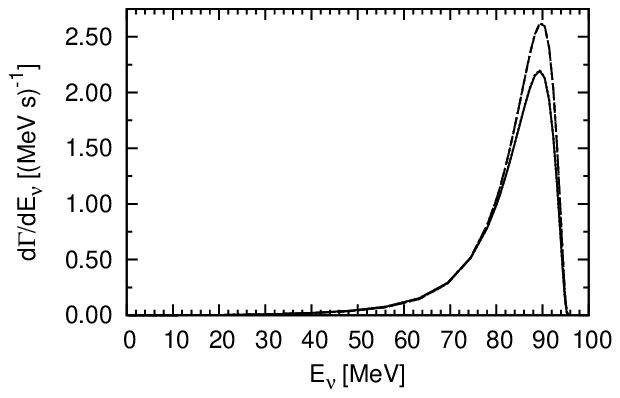}
\end{center}
\caption{The 3N force effects in the differential capture rates 
    ($F=0$) $ {d\Gamma }^{F=0}/ {dE_\nu} $ (top row),
    ($F=1$) $ {d\Gamma }^{F=1}/ {dE_\nu}   $ (middle row)
and (total) $ {d\Gamma }/ {dE_\nu} $ (bottom row)
calculated without the 3N force (dash-dotted), including the 3N force only in the initial state 
(dotted line), only in the final state (dashed) and with the 3N forces taken consistently 
in the initial and final states (solid line).
The dash-dotted and dashed lines overlap. The same is true for the dotted 
and solid lines.
\label{fig3}}
\end{figure}

\begin{figure}
\begin{center}
\includegraphics[width=7cm]{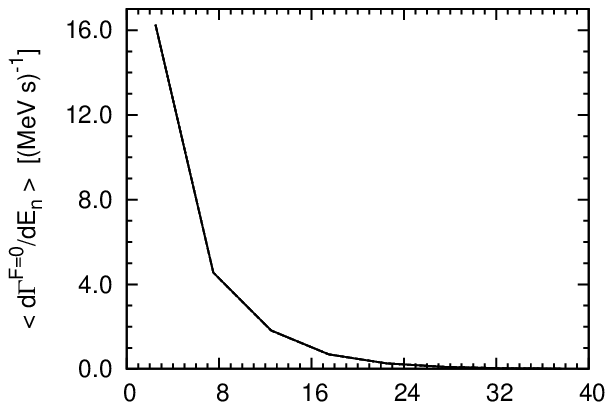}
\includegraphics[width=7cm]{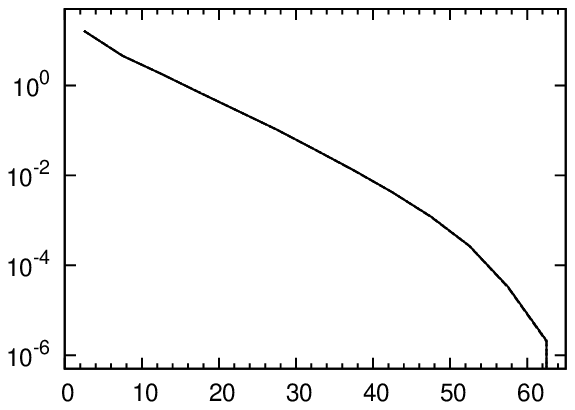}
\end{center}
\begin{center}
\includegraphics[width=7cm]{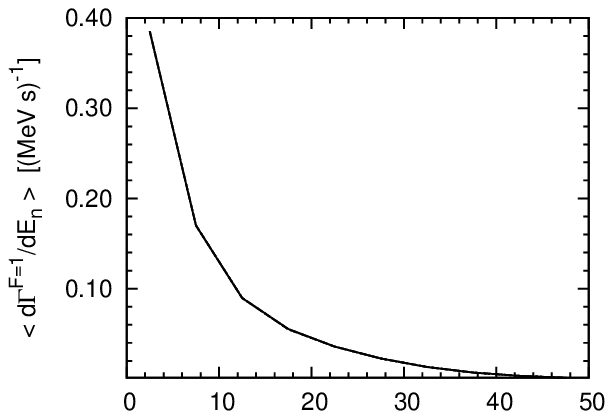}
\includegraphics[width=7cm]{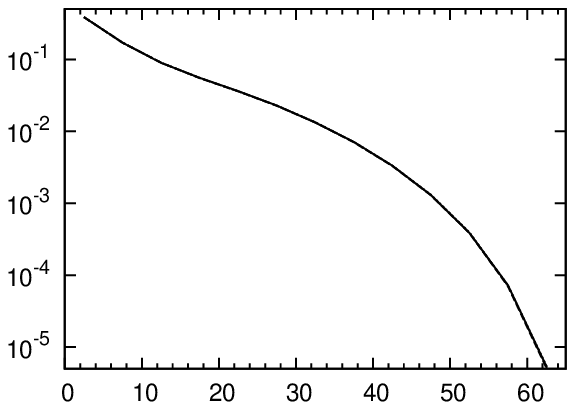}
\end{center}
\begin{center}
\includegraphics[width=7cm]{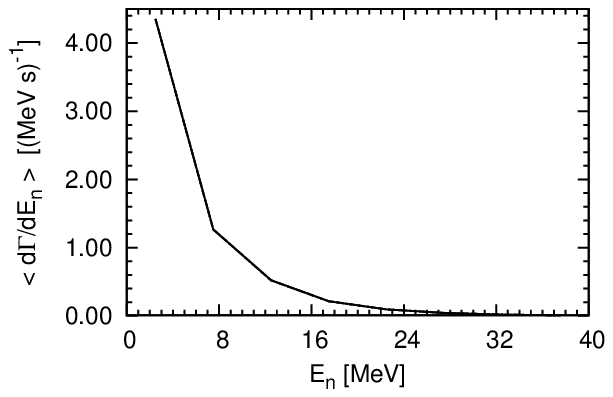}
\includegraphics[width=7cm]{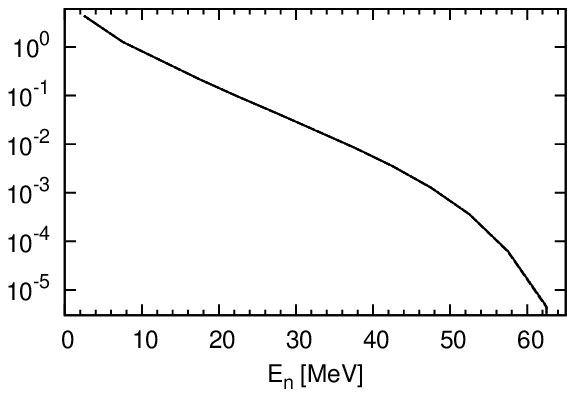}
\end{center}
\caption{The differential capture rates
($F=0$) $ \langle  {d\Gamma }^{F=0}/ {dE_n} \rangle $ (top row),
($F=1$) $ \langle  {d\Gamma }^{F=1}/ {dE_n} \rangle $ (middle row)
and (total) $ \langle  {d\Gamma }/ {dE_n} \rangle $ (bottom row),   
for the $ \mu^- + ^{3}\textrm{H} \rightarrow \nu_\mu + n + n + n$
process averaged over 5~MeV neutron energy bins. The same results 
are shown on a linear (left panel) and a logarithmic (right panel) scale. 
The predictions are obtained using the full solution of Eq.~(\ref{u}).
The three overlapping curves represent results, where the energy 
of nucleon~1 (solid line), nucleon~2 (dashed line) and nucleon~3 (dotted line) 
is considered.
\label{fig4}}
\end{figure}

\begin{figure}
\begin{center}
\includegraphics[width=7cm]{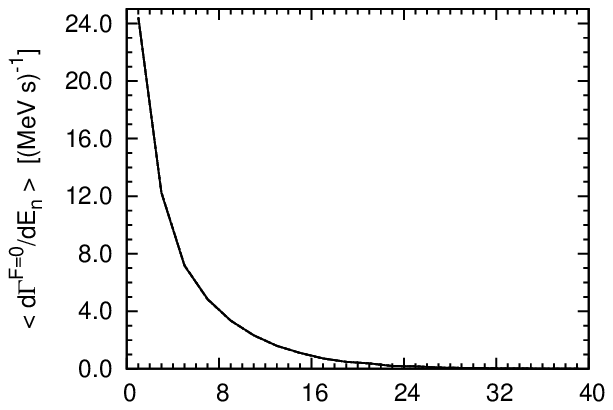}
\includegraphics[width=7cm]{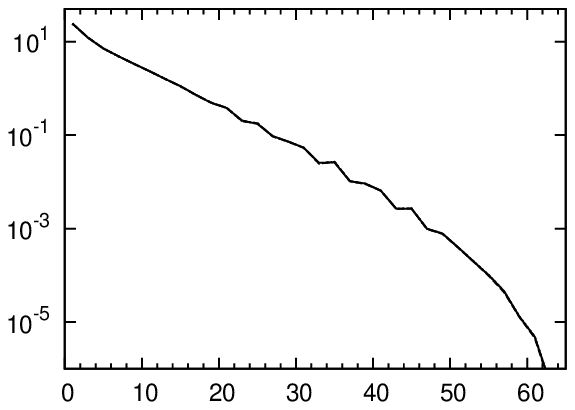}
\end{center}
\begin{center}
\includegraphics[width=7cm]{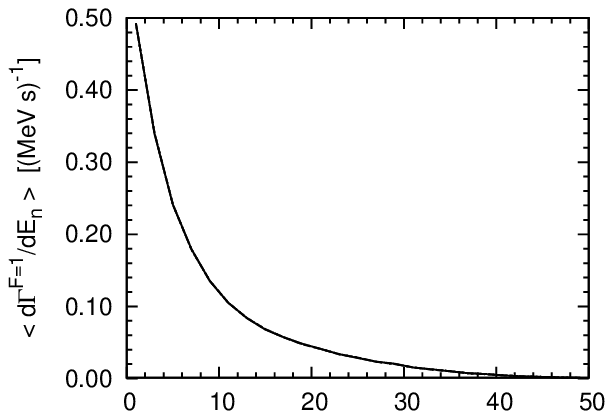}
\includegraphics[width=7cm]{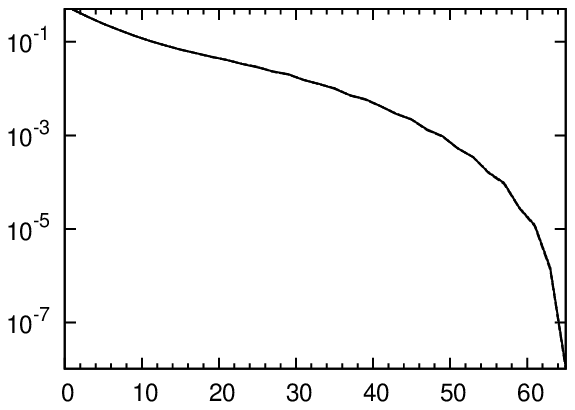}
\end{center}
\begin{center}
\includegraphics[width=7cm]{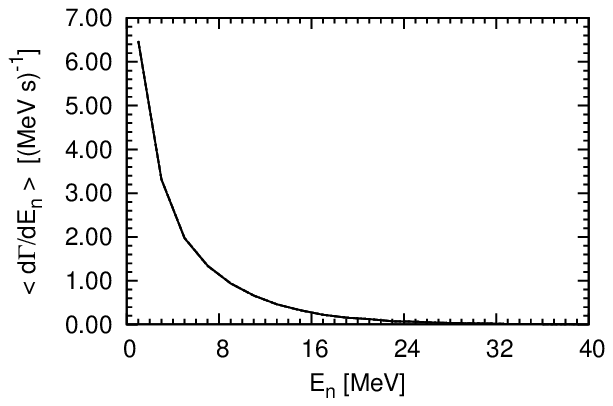}
\includegraphics[width=7cm]{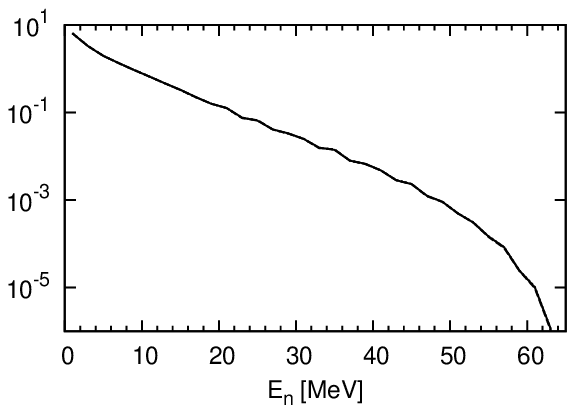}
\end{center}
\caption{The same as in Fig.~\ref{fig4} but the capture rates
are averaged over 2~MeV neutron energy bins. 
\label{fig5}}
\end{figure}

We start the discussion of our predictions with Fig.~\ref{fig1},
where the differential capture rates 
$d \Gamma/ dE_{\nu}$
are shown for the considered 
$ \mu^- + ^3\textrm{H} \rightarrow \nu_\mu + n + n + n$
capture reaction and
effects of the relativistic corrections in the single nucleon current 
operator are studied. The results are calculated with the 2N forces only.
Although the rates are not independent 
(see Eq.~(\ref{notindpd})), we display all the three quantities. Clearly, 
the values for $F=0$ are much bigger than for $F=1$. Since the $F=0$ rate 
dominates, the $F=0$ (top row) and total (bottom row) rates change 
with the neutrino energy in a very similar way. 
They rise very slowly for small neutrino energies and
show a strong maximum in the vicinity
of the maximal neutrino energy, where the phase space factor 
reduces all the differential rates to zero.
The behavior of the $F=1$ rate is different. Its values grow much faster 
with the neutrino energy and the corresponding maximum is therefore very broad, 
reaching quite small neutrino energies. The relativistic effects are hardly 
visible on the linear scale, except for the very peak area, where two 
curves stop overlapping. The relativistic effects 
reduce the maximal values of the $F=0$ rate (by approximately 4 \%)
and the total rate (by approximately 2~\%) and increase the
value of the $F=1$ rate (by nearly 9~\%).
The changes of the total (integrated) rates are discussed below.

In Fig.~\ref{fig2} the same differential rates are shown but 
they are calculated with three different types of 3N dynamics. 
We display predictions obtained using the plane wave (PW) approximation
(see Eq.~(\ref{nnn3pw})), results of the calculations that employ
only 2N forces (given by the AV18 potentials~\cite{av18}) 
to calculate the initial as well as final 3N states, 
and finally predictions based on a consistent
treatment of the initial and final states, taking additionally a
3N force (the Urbana~IX \cite{urbana} potential) into account.
Final state interactions are very important. They not only 
change the PW predictions by a factor of $2$ but alter also the 
shapes of the curves and their peak positions. We thus confirm the findings 
of Refs.~\cite{torre1979,dzhibuti1984} obtained with completely 
different frameworks and much simpler forces. Like in Ref.~\cite{golak2014},
we also study the 3N force effects. They are clearly visible in the peak 
areas, where the predictions including the 3N force drop 
by approximately 20~\%. 
These peak reductions
are quite similar to the two-body and three-body
break-up cases studied in Ref.~\cite{golak2014} for muon capture on $^3$He.
In these calculations the same single nucleon current operator
including relativistic corrections is used. Note that the PW results 
are obtained with the initial 3N bound state calculated solely with the 2N 
forces.

It is interesting to trace back the origin of such large 
3N force effects. To this end
in Fig.~\ref{fig3} we display results of four different calculations 
with the same single nucleon current operator. 
We can neglect the 3N force both in the 3N bound and scattering states, 
include it only in the initial state, only in the final state and finally
use it consistently in the 3N bound state and in the 3N continuum.
For all the three differential rates 3N force contributions
to the final three-neutron scattering state are very small and do not reach 
even 1~\%. 
Only the inclusion of the 3N force in the initial bound state
is decisive for the calculations.
Therefore in Fig.~\ref{fig3} we see two groups of curves, each obtained
with the same initial bound state.

It is clear that quantities like the differential capture rates 
$d \Gamma/ dE_{\nu}$
would be extremely hard to measure. More realistic is to expect 
that the capture rates
$d \Gamma/ dE_{n}$ ($E_n$ is the final neutron energy)
will be accessed experimentally. Such measured rates will be in practice
averaged over certain intervals of the neutron energies.
In order to calculate the corresponding 
theoretical rates we do not introduce 
any dedicated kinematics but use the 
same steps as required to obtain the total rates 
according to Eq.~(\ref{gnnn}).
Thus we are sure that the calculations
of the (averaged) differential rates $ \langle  {d\Gamma }/ {dE_n} \rangle $
are consistent with 
the calculation of the total rate $\Gamma$, where
we obtain first the capture rates
$ {d\Gamma }/ {dE_\nu} $
at 36 values of the final neutrino energy,
solving for each of them the corresponding
Faddeev-like equation (\ref{u}).
These neutrino energies are distributed 
with special emphasis on the region
in the vicinity of the maximal neutrino energy.
Therefore we use the same
formulas and codes as for the total
${\Gamma }$
capture rate, performing integrals over the whole phase
space. However, the contribution to a given neutron energy interval 
comes only from the integrand with a proper kinematical ``signature''
\cite{golak2014}.

This kinematical ``signature'' is easy to obtain
because the individual momenta of the three outgoing neutrons
can be evaluated from Eqs.~(\ref{pq})
\begin{eqnarray}
{\bf p}_1 = -\frac13 {\bf p}_\nu  + {\bf q} \, , \nonumber \\
{\bf p}_2 = -\frac13 {\bf p}_\nu  + {\bf p} - \frac12 {\bf q} \, , \nonumber \\
{\bf p}_3 = -\frac13 {\bf p}_\nu  - {\bf p} - \frac12 {\bf q} \, .
\label{e1e2e3}
\end{eqnarray}
Since the outgoing neutrons are identical, we have 
actually three possibilities to calculate the final neutron energy:
$E_n = \frac1{2 M} {\bf p}_i^2 $, $i=1,2,3$. This can be used as a 
nontrivial test of the final state antisymmetrization.

In Fig.~\ref{fig4} 
the capture rates:
$ \langle  {d\Gamma }^{F=0}/ {dE_n} \rangle $,
$ \langle  {d\Gamma }^{F=1}/ {dE_n} \rangle $ 
and 
$ \langle  {d\Gamma }/ {dE_n} \rangle $,
averaged over 5~MeV neutron energy bins are shown 
both on a linear and a logarithmic scale. The reason for that is 
that the rates change by several  orders of magnitude in the allowed 
interval of the neutron energy. These results are obtained with the full
inclusion of the 3N force. The three curves denote the results,
where the energy of nucleon~1, nucleon~2 and nucleon~3 is taken
as the neutron energy $E_n$. The three lines completely overlap,
which confirms the proper antisymmetrization of the final three-neutron states.
In these calculations we use 
thus 36 $E_\nu$ points, 36 $\theta_{q}$ points,
36 $\theta_{p}$ points,
36 $\phi_{p}$ points
and also 36 values of the magnitude of the relative ${\bf p}$ momentum.
Note that due to the rotational invariance of the unpolarized rate, we may
choose $\phi_{q}=0$. These numbers of points are sufficient to get 
smooth curves for the 5~MeV neutron energy bins. 
In Fig.~\ref{fig5} we show the same capture rates 
as in Fig.~\ref{fig4} 
but now they are averaged over smaller 2~MeV neutron energy bins. From the wavy character of some lines (visible
on the logarithmic scale) one can infer that an average over smaller than 5~MeV energy
intervals requires a finer grid of $E_\nu$ points.

We supplement the results presented in Figs.~\ref{fig1}--\ref{fig5}
by giving the corresponding values of integrated capture rates
in Table~\ref{tabbr}, together with earlier
theoretical predictions of Refs.~\cite{phillips1975,torre1978,torre1979,dzhibuti1984}.
From the first two rows of this table it is clear that final state interactions 
taken in the form of 2N forces enhance the plane wave results (given in brackets) 
by 34--38~\%. This effect is similar for the nonrelativistic single nucleon 
current operator and for the current operator containing relativistic corrections.
The relativistic corrections reduce the $F=0$ rate by approximately 3.4~\% 
and raise the $F=1$ by more than 6~\%. The effect on the total rate is weaker: this 
rate is reduced by approximately 2.6~\%. (All these changes are discussed for the ``Full 2NF''
results.)
Note that this effect is slightly larger than 
    for the $ \mu^- + ^3\textrm{He} \rightarrow \nu_\mu + ^3\textrm{H}$ process,
    for which the total capture rate, calculated with the
    AV18 2N potential~\cite{av18} augmented
    by the Urbana~IX 3N force~\cite{urbana},
    is reduced by 1.6~\%, when the relativistic
    corrections are included in the single-nucleon current.
    This information was already obtained by one of the authors (L.E.M.), 
    when the calculations published in Ref.~\cite{Mar02} were performed,
    but it was not included in the publication. 
The inclusion of the 3N force decreases all the three rates.
This reduction is stronger for the $F=0$ and total rates (approximately 12~\%)
than for the $F=1$ one (approximately 7.5~\%).
The reduction due to the 3N force is a common feature of the
    muon capture on the $A=3$ systems. Already in Ref.~\cite{Mar02} it was shown
    that there is a significant correlation between the total capure rate
    for the $ \mu^- + ^3\textrm{He} \rightarrow \nu_\mu + ^3\textrm{H}$ reaction
    and the $A=3$ binding energies.

It is very interesting to notice that much earlier theoretical 
predictions agree quite well with our new results. This is true not only 
for the plane wave results but also for the calculations that consistently 
used 2N forces in the initial and final 3N states. 

Results of our most advanced approach (the AV18 nucleon-nucleon potential 
augmented by the Urbana~IX 3N force and the single nucleon current operator
containing relativistic corrections) are given in 
the third row of Table~\ref{tabbr} and read  
$\Gamma^{F=0} = 118.7\, {\rm s}^{-1} $,
$\Gamma^{F=1} = 3.92\, {\rm s}^{-1} $ and
$\Gamma = 32.6\, {\rm s}^{-1} $.

\section{Summary and conclusions}
\label{section4}
This paper constitutes an important step towards a consistent framework 
for calculations of all muon capture processes on
the deuteron and $A=3$ nuclei. 
This requires that the initial and final nuclear states 
are calculated with the same Hamiltonian and that the weak current operator 
is fully consistent with the nuclear forces. 
Results of such calculations can be then compared with precise 
experimental data to improve our understanding of muon capture and other
weak reactions. 

In the present paper we study the
$\mu^- + ^3\textrm{H} \rightarrow \nu_\mu + n + n + n$
process 
in the framework close to the potential model
approach of Ref.~\cite{prc83.014002} but
with the single nucleon current operator.
This is a continuation of our work 
from Ref.~\cite{golak2014}, where other 
capture reactions:
$\mu^- + ^2\textrm{H} \rightarrow \nu_\mu + n + n$,
$\mu^- + ^3\textrm{He} \rightarrow \nu_\mu + ^3\textrm{H}$,
$\mu^- + ^3\textrm{He} \rightarrow \nu_\mu + n + d$
and
$\mu^- + ^3\textrm{He} \rightarrow \nu_\mu + n + n + p$
were described in the same momentum space framework.
We use the results of Ref.~\cite{golak2014}
on the partial wave decomposition of the single 
nucleon current operator, the number of partial wave states 
necessary to reach convergence of the results and the 
simple method to obtain the averaged capture rates 
from the calculations of the total rate. 
It is quite understandable that also for this reaction 
the nonrelativistic kinematics can be safely used.

Using for the first time modern semi-phenomenological 2N and 3N 
forces, we give predictions for the differential 
$d\Gamma / dE_{\nu}$ 
capture rates as well as for the corresponding integrated capture rates
$\Gamma$ 
and the averaged 
$ \langle d\Gamma / dE_{n} \rangle$ 
capture rates, taking additionally into account the $F=0$ and $F=1$ 
hyperfine states of the muon-tritium atom.
Our best numbers (from the calculations employing 
the AV18 2N potential and the Urbana~IX 3N force and using the single nucleon current operator
containing relativistic corrections) are: 
$\Gamma^{F=0} = 118.7\, {\rm s}^{-1} $,
$\Gamma^{F=1} = 3.92\, {\rm s}^{-1} $ and
$\Gamma = 32.6\, {\rm s}^{-1} $.

Our predictions obtained with the 2N force alone
are in a rather good agreement with much older theoretical 
predictions from 
Refs.~\cite{phillips1975,torre1978,torre1979,dzhibuti1984}. 
Our results cannot be confronted
with experimental data at the moment. It is clear that a measurement
of the reaction considered in this paper would be extremely difficult.
However, due to the presence of three neutrons in the final state 
and their two- and three-body interactions,  
theoretical and experimental investigations of this reaction 
are very interesting and important.

We are well aware that 
the full understanding of the muon capture processes requires
the inclusion of at least 2N 
contributions to the nuclear current operators. First steps
in this direction were made in Ref.~\cite{golak2014}.
We do hope that, even in the present shape, our predictions will serve 
as an important benchmark.
In the near future we plan to perform 
more complete calculations using  
the locally regularized chiral
2N potential~\cite{epel1,epel2}, supplemented by the 
consistently regularized chiral 3N forces~\cite{epel2002,heb1}
and electroweak current operators~\cite{epel3}.

\acknowledgments
This study was supported by the Polish National Science Center under 
Grants No. DEC-2013/10/M/ST2/00420 and additionally DEC-2013/11/N/ST2/03733.
The numerical calculations have been performed on the supercomputer clusters of the JSC, J\"ulich,
Germany.

\end{document}